# The existence of a quantum phase transition in a Hubbard model on a square lattice.

Valentin Voroshilov

Physics Department, Boston University, Boston, MA, 02215, USA

A novel canonical transformation is offered as the mean for studying properties of a system of strongly correlated electrons. As an example of the utility of the transformation, it is used to demonstrate the existence of a quantum phase transition in a Hubbard model on a square lattice. An Appendix presents two cases with a negative result.

74.20.Mn    71.10.Li

The Hubbard model[1] is a model of choice to study properties of a system of strongly interacting electrons. A recent paper[2] offers a novel canonical transformation as a tool for analyzing the properties of the model. However, the approach had two major limitations: the term describing on-site interaction had been omitted from the Hamiltonian; the transformation did not provide a crossover between a "normal" and an "anomalous" state.

This paper offers an approach without the mentioned limitations, as well as shows a possible generalization to a 2-dimensional lattice.

The major simplification we use is treating a $N$x$N$ lattice as a composition of four independent $N/2$x$N/2$ lattices. Hamiltonian (1) below is written for one of the sublattices using the standard symbolism; it is a two-band Hamiltonian which includes on-site interaction and nearest-neighbors interaction terms; Hamiltonians for all four sublattices are identical (besides the numeration of the sites). In (1), for Fermi operators $a^+_{\alpha ij\sigma}$ and $a_{\alpha ij\sigma}$, $\alpha = 1,2$ numerates the bands, $i$ and $j$ represent $x$ and $y$ coordinates of a site (for the first sublattice $i, j = 1,...\frac{N}{2}$, etc.), $\sigma = \pm$ represent spin components of an electron.

Valentin Voroshilov, valbu@bu.edu, Physics Department, Boston University, 590 Commonwealth Ave., Boston, MA 02215

$$H = -t \sum_{\substack{<\vec{r}, \vec{r}'> \\ \alpha,\beta,\sigma}} a^+_{\alpha\vec{r}\sigma} a_{\beta\vec{r}'\sigma} + U \sum_{\vec{r},\alpha,\beta} n_{\alpha\vec{r}+} n_{\beta\vec{r}-} + U \sum_{\vec{r},\sigma} n_{1\vec{r}\sigma} n_{2\vec{r}\sigma} + V \sum_{\substack{<\vec{r}, \vec{r}'> \\ \alpha,\beta}} n_{\alpha\vec{r}+} n_{\beta\vec{r}'-} + W \sum_{\substack{<\vec{r}, \vec{r}'> \\ \alpha,\beta}} n_{\alpha\vec{r}+} n_{\beta\vec{r}'+}. \quad (1)$$

The novel canonical transformation is a composition of two canonical transformations: the first transformation happens in the real space and involves electrons from two neighboring sites from the opposite rows, and the second one makes a transition to the momentum space of the system.

The first transformation introduces a set of new Fermi operators $b_{ik}, b^+_{ik}$ ($k = 1,2,3,4$); the new operators combine the creation and annihilation operators for electrons sitting opposite each other on two neighboring rows; the transformation can be written in the following form (the example is for one sublattice):

$$\begin{pmatrix} a_{\alpha ij+} \\ a_{\alpha ij-} \\ a_{\alpha ij+1+} \\ a_{\alpha ij+1-} \end{pmatrix} = B \cdot \begin{pmatrix} b_{\alpha ij1} \\ b_{\alpha ij2} \\ b_{\alpha ij3} \\ b_{\alpha ij3} \end{pmatrix} + D \cdot \begin{pmatrix} b^+_{\alpha ij1} \\ b^+_{\alpha ij2} \\ b^+_{\alpha ij3} \\ b^+_{\alpha ij4} \end{pmatrix}; \quad B = \begin{pmatrix} 1-q & -q & -q & q \\ q & 1-q & q & q \\ q & -q & 1-q & -q \\ -q & -q & q & 1-q \end{pmatrix};$$

(2)

$$D = \sqrt{\frac{2q(1-2q)}{1-2q+4q^2}} \cdot \begin{pmatrix} -q & -q & q & 1-q \\ q & -q & 1-q & -q \\ -q & -1+q & -q & -q \\ -1+q & q & q & -q \end{pmatrix}; \quad 0 < q < 0.5.$$

Together with four equations for $a^+_{\alpha ij\sigma}$ relations (2) provide a canonical transformation. The solution with $q = 0$ leads to a one-to-one correspondence between $\{a, a^+\}$ and $\{b, b^+\}$ operators, hence transformation (2) preserves the crossover between a "normal" and an "anomalous" states.

Transformation (2) has been designed specifically to test properties related to propagation of *correlated* spin waves (similarly to the Bogolubov[3] transformation which had been designed to test the properties related to the existence of bound electron pairs).

The second transformation is a standard transition from operators acting in a real space to operators acting in a momentum space (for further calculations Planck's constant and a lattice constant are set to unity):

$$b_{\alpha l j k} = \frac{1}{\sqrt{\frac{N}{2}}} \sum_{p_x \in \Omega_\alpha} e^{ip_x l} \frac{1}{\sqrt{\frac{N}{4}}} \sum_{p_y \in \Omega_\alpha} e^{ip_y j} b_{p_x p_y k}, \quad k = 1, 2, 3, 4. \tag{3}$$

The values for momenta are confined by the following sets: for the first band, $p_x, p_y \in \Omega_1$, $p_x = \pm \frac{4\pi n}{N}, n = 0,1,...,\frac{N}{4}$, $p_y = \pm \frac{8\pi n}{N}, n = 0,1,...,\frac{N}{8}$; for the second band, $p_x, p_y \in \Omega_2$, $p_x = \pm(\pi + \frac{4\pi n}{N}), n = 0,1,...,\frac{N}{4}$, $p_y = \pm \frac{8\pi n}{N}, n = 0,1,...,\frac{N}{8}$; as it is seen the first band is filled. Following the notion of waves traveling in one direction (namely, in the x-direction for the first sublattice) for the second band we assume that only one component of the momentum ($p_x$) differs the second band from the first one (note: if waves travel in x-direction, sites have to be counted in that direction but only *pairs* of rows are to be counted in the orthogonal y-direction).

To estimate the ground state energy of the system we use a well known variational approach[4]. First, we define a test ground state vector in the following form:

$|E_0> = \prod_{\substack{|p_x|<\Pi \\ |p_y|<\pi}} b^+_{p_x p_y 1} b^+_{p_x p_y 2} b^+_{p_x p_y 3} b^+_{p_x p_y 4} |0>$, with $|0>$ to be the vacuum for operators $b_{p_x p_y k}$, i.e. $b_{p_x p_y k} |0> = 0$ (note: contrary to BCS[5] theory, we do not presume any electron paring in the momentum space); for a two-band Hamiltonian $\pi < \Pi < 2\pi$.

For the first sublattice we apply transformation (2) to pairs of rows parallel to x-direction, i.e. i-index numerates the sites, $i = 1, ..., N/2$ and j-index numerates the pairs of rows, $j = 2k-1$ with $k = 1, ..., N/4$. In this case transformation (2) reflects correlation between waves traveling in the rows with $j = 1$ and $j = 2$, then with $j = 3$ and $j = 4$, then with $j = 5$ and $j = 6$, etc.

For the second sublattice we apply transformation (2) again to pairs of rows parallel to x-direction, i.e. i-index numerates the sites, $i = 1, ..., N/2$ and j-index numerates the pairs of rows, $j = 2k$ with $k = 1, ..., N/4$, i.e. now transformation (2) reflects correlation between waves traveling in the rows with $j = 2$ and $j = 3$, then with $j = 4$ and $j = 5$, then with $j = 6$ and $j = 7$, etc.

For the third and the fourth sublattices we apply transformation (2) to pairs of rows parallel to y-direction, i.e. we "switching" x and y directions to preserve the equivalence for these directions for the whole lattice (it is clear that transformation (2) brakes such an equivalence). Calculating the ground state energy expectation value for each sublattice, however, leads to the same result for all four.

Transformation (2) does not conserve the number of particles in the system, hence, we demand that the expectation value of the operator for the total number of electrons is equal to the actual number of electrons in the subsystem, $N_e/4$ ($N_e$ is the total number of electrons on the whole lattice);

$$N_e/4 = <E_0 | \sum_{\alpha ij\sigma} n_{\alpha ij\sigma} | E_0>, \qquad 2 < \frac{N_e}{N^2} < 4. \qquad (4)$$

Because $\pi < \Pi < 2\pi$, we impose an additional condition:

$$0 < \frac{\frac{N_e}{2N^2} - 1}{1 - 4q + 8q^2} < 1. \qquad (5)$$

When $q \neq 0$ (or $q \neq 0.5$) an anomalous electron pair correlation function $<E_0 | a_{ij+} a_{ij+1-} | E_0>$ is not zero (assuming $j$ and $j+1$ describe the pair of rows connected by the canonical transformation). This is a sign of a new phase in the system (it is natural to call this phase as "anomalous"). This new phase can be reached only when parameters of the Hamiltonian satisfy the given conditions and the ground state energy of the system reaches its minimum at $q \neq 0$ or $q \neq 0.5$ ($0 < q < 0.5$).

Calculations for a one-band Hamiltonian did not lead to the existence of an "anomalous" state for all values of the parameters of the Hamiltonian.

Calculations for a two-band Hamiltonian with $V = W = 0$ (the model with only the on-site interaction) did not lead to the existence of an "anomalous" state.

Calculations for a two-band Hamiltonian have lead to the existence of an "anomalous" state for certain region of parameters, for example $<E_0 | a_{ij+} a_{ij+1-} | E_0> \neq 0$ when $\frac{N_e}{N^2} = 3.67$, $t/U = 0.2$, $V/U = 0.7$, and $W = 0$.

This is a promising result showing the existence of a transition between a "normal" and "anomalous" phases when parameters of the system change.

It is worth to note that the test ground state vector and the Hamiltonian are not based on an assumption of an existence of an effective attraction between electrons[6].

There is methodical task, however, which must be done before drawing a phase diagram of the system or extending the approach to higher temperatures (which is a fairly straightforward task).

It is clear that the matrices used in transformation (2) are not the only one which can be used for a canonical transformation of the type. A canonical transformation between $\{a, a^+\}$ and $\{b, b^+\}$ operators can be described with the means of an 8x8 matrix $M$; $M = \begin{pmatrix} B D \\ D B \end{pmatrix}$, with $B$ and $D$ are 4x4 matrices which satisfy two conditions: $B \cdot B^T + D \cdot D^T = 1$, $B \cdot D^T + D \cdot B^T = 0$. The set of such matrices $M$ forms a subgroup of the SO(8) group (does it mean that the Hubbard model is as rich as Supergravity?). The minimization of the ground state energy based on the use of the complete subgroup is a part of the ongoing investigation.

**Appendix**

Negative result might also have an interest, presenting what might be as a dead end search.
The Appendix presents two cases when an "anomalous" phase had not been found (it is not clear yet if this is the feature of the transformation (of the specific representation), or of the probe ground state vector, or of the Hamiltonian).
The simplest version of a two-band Hubbard model on a square lattice had been considered

$$H = - \sum_{\substack{<\vec{r},\vec{r}'> \\ \alpha,\beta,\sigma}} t_\alpha a^+_{\alpha\vec{r}\sigma} a_{\beta\vec{r}'\sigma} + U \sum_{\vec{r},\alpha,\beta} n_{\alpha\vec{r}+} n_{\beta\vec{r}-} + W \sum_{\vec{r},\sigma} n_{1\vec{r}\sigma} n_{2\vec{r}\sigma} . \tag{6}$$

A transformation had been applied to Hamiltonian (6), which combines only electrons from the same band and can be described with the means of a two-parametric 2x2 matrices

$$\begin{pmatrix} a_{\alpha\vec{r}+} \\ a_{\alpha\vec{r}-} \end{pmatrix} = B \cdot \begin{pmatrix} b_{\alpha\vec{r}1} \\ b_{\alpha\vec{r}2} \end{pmatrix} + D \cdot \begin{pmatrix} b^+_{\alpha\vec{r}1} \\ b^+_{\alpha\vec{r}2} \end{pmatrix}; \quad B = \begin{pmatrix} 1-v & z \\ -z & 1-v \end{pmatrix} \; ; \; D = \begin{pmatrix} bz & -b(1-v) \\ b(1-v) & bz \end{pmatrix}, \tag{7}$$

where $b = a / \sqrt{1 - 2v + v^2 + z^2}$, and $\sqrt{2} z = \sqrt{-1 + \sqrt{1 - 4a^2} + 4v - 2v^2}$ (the parameterization leads to two independent parameters, namely $a$ and $v$). The second part of the transformation is a standard transition from operators acting in a real space to operators acting in a momentum space:

$$b_{\alpha\vec{r}k} = \frac{1}{N} \sum e^{i\vec{p}\vec{r}} b_{\alpha\vec{p}k} , \qquad \alpha, k = 1, 2. \tag{8}$$

A probe ground state vector is defined in the following form:

$|E_0> = \prod_{\substack{|p_x|<\pi \\ |p_y|<\pi}} \prod_{\substack{|q_x|<\varepsilon \\ |q_y|<\varepsilon}} b^+_{Ip_xp_y1} b^+_{Ip_xp_y2} b^+_{IIq_xq_y1} b^+_{IIq_xq_y2} |0>$, with $|0>$ to be the vacuum for operators $b_{\alpha p_x p_y k}$, i.e.

$b_{\alpha p_x p_y k}|0>=0$. We also demand that the expectation value of the operator for the total number of electrons is equal to the actual number of electrons in the system, $N_e$;

$$N_e = <E_0| \sum_{\alpha \bar{r} \sigma} n_{\alpha \bar{r} \sigma} |E_0>. \tag{9}$$

In general $0 < n = \frac{N_e}{N^2} < 4$, however, with our choice of the probe ground state vector Eq. (9) leads to $n = 2(1 + m^2\sqrt{1-4a^2})$, where $-0.5 < a < 0.5$; $2 < n < 4$; $m = <b^+_{\alpha \bar{p} k} b_{\alpha \bar{p} k}> = \sqrt{(n/2-1)/\sqrt{1-4a^2}}$. Further analysis had not shown the existence of an "anomalous" phase.

Another transformation which combines all eight operators related to the same site had been applied to Hamiltonian (6), which can be written in the following matrix form (expression for creation operators is omitted)

$$\begin{pmatrix} a_{I\bar{r}+} \\ a_{I\bar{r}-} \\ a_{II\bar{r}+} \\ a_{II\bar{r}-} \end{pmatrix} = B \cdot \begin{pmatrix} b_{\bar{r}1} \\ b_{\bar{r}2} \\ b_{\bar{r}3} \\ b_{\bar{r}3} \end{pmatrix} + D \cdot \begin{pmatrix} b^+_{\bar{r}1} \\ b^+_{\bar{r}2} \\ b^+_{\bar{r}3} \\ b^+_{\bar{r}4} \end{pmatrix}; \qquad b_{\bar{r}k} = \frac{1}{N} \sum e^{i\bar{p}\bar{r}} b_{\bar{p}k}, \quad k=1,2,3,4. \tag{11}$$

A probe ground state vector is defined as: $|E_0> = \prod_{\substack{|p_x|<\varepsilon \\ |p_y|<\varepsilon}} b^+_{\bar{p}1} b^+_{\bar{p}2} b^+_{\bar{p}3} b^+_{\bar{p}4} |0>$, with $|0>$ to be the vacuum for operators $b_{\bar{p}k}$, i.e. $b_{\bar{p}k}|0>=0$; $N_e = <E_0| \sum_{\alpha \bar{r} \sigma} n_{\alpha \bar{r} \sigma} |E_0>$

Matrices $B$ and $D$ satisfy two conditions: $B \cdot B^T + D \cdot D^T = 1$, $B \cdot D^T + D \cdot B^T = 0$; a specific representation with six independent parameters was used:

$$B = \begin{pmatrix} 1-q & z & -y & x \\ -z & 1-q & x & y \\ y & -x & 1-q & z \\ -x & -y & -z & 1-q \end{pmatrix}, \quad D = -A*(B^T)^{-1}, \quad A = \begin{pmatrix} 0 & a & b & c \\ -a & 0 & c & -b \\ -b & -c & 0 & a \\ -c & b & -a & 0 \end{pmatrix}, \tag{12}$$

with $q = 1 - \frac{\sqrt{2}}{2}\sqrt{1-2(x^2+y^2+z^2) - \sqrt{1-4(a^2+b^2+c^2)}}$.

Further analysis had not shown the existence of an "anomalous" phase.

It is noteworthy to point out that the presence of a set of canonical transformations applicable to the Hamiltonian could be seen as a manifestation of the existence in the system of competing symmetries[7].